\documentclass[superscriptaddress,aps,pra,onecolumn, showpacs,nofootinbib,longbibliography,notitlepage]{revtex4-2}

\usepackage{pgfplots}
\pgfplotsset{compat=1.18}
\usepackage[utf8]{inputenc}
\usepackage[T1]{fontenc}     
\usepackage[british]{babel}  
\usepackage[sc,osf]{mathpazo}
\usepackage[scaled=0.86]{berasans}  
\usepackage[colorlinks=true, allcolors=blue, urlcolor=blue]{hyperref}  
\usepackage{graphicx} 
\usepackage[babel]{microtype}  
\usepackage{amsmath,amssymb,amsthm,bm,amsfonts,mathrsfs,bbm} 

\usepackage{xspace}  
\usepackage{pgfplots}
\usepackage{xcolor,colortbl}
\usepackage{array}
\usepackage{bigstrut}\usepackage[normalem]{ulem}
\usepackage{mathtools}
\usepackage{physics}

\usepackage{caption}

\usepackage{subcaption}
\usepackage{tabularx}
\newcolumntype{C}{>{\centering\arraybackslash}X}
\usepackage{lipsum}
\usepackage[inkscapelatex=true]{svg}
\usepackage{hyperref}
\hypersetup{
	colorlinks=true,
	linkcolor=red,
	filecolor=blue,      
	urlcolor=blue,
	citecolor=purple,
}

\newcounter{subeq}

\newcommand{\be}{\begin{equation}}
	\newcommand{\ee}{\end{equation}}
\newcommand{\bea}{\begin{eqnarray}}
	\newcommand{\eea}{\end{eqnarray}}
\newcommand{\bes}{\begin{equation*}}
	\newcommand{\ees}{\end{equation*}}
\newcommand{\beas}{\begin{eqnarray*}}
	\newcommand{\eeas}{\end{eqnarray*}}
\newcommand{\proj}[1]{\ket{#1}\bra{#1}}

\def\tr{\mathrm{tr}}

\newtheorem*{thm*}{Theorem}

\newtheorem*{lem*}{Lemma}

\newtheorem*{prop*}{Proposition}

\newtheorem*{lipschitzLem*}{Lemma \ref{lipschitz}}
\newtheorem*{lipschitzCubeLem*}{Lemma \ref{lipschitzCube}}
\newtheorem*{pgmNearlyOptimalThm*}{Theorem \ref{pgmNearlyOptimal}}

\begin{document}

\title{Multipartite Hardy paradox unlocks device-independent key sharing}

\author{Ranendu Adhikary}
 \email{ronjumath@gmail.com}
\affiliation{Cryptology and Security Research Unit, Indian Statistical Institute, 203 B.T. Road, Kolkata 700108, India}

\author{Mriganka Mandal}
 \email{mriganka@isical.ac.in}
\affiliation{Cryptology and Security Research Unit, Indian Statistical Institute, 203 B.T. Road, Kolkata 700108, India}

\begin{abstract}
We introduce a device-independent quantum key distribution protocol for $N$ parties, leveraging the multipartite Hardy paradox. Unlike traditional multipartite protocols that extract the key from measurement outcomes, our method generates the shared secret key directly from the parties’ measurement settings, achieving a positive key rate certified by the paradox’s maximal violation. Notably, the paradox enables any two parties to create a secret key with a rate much higher than the $N$-party key, due to more robust pairwise correlations. This unique capability, inherent to the multipartite Hardy paradox, allows for tailored key distribution within the group, enhancing flexibility. Using non-maximally entangled states, our protocol establishes a new paradigm for device-independent conference key agreement, ensuring information-theoretic security in untrusted quantum networks. 
\end{abstract}
\maketitle

\section{Introduction}

Secure communication among multiple parties underpins applications ranging from diplomatic exchanges to distributed quantum computing~\citep{Kimble2008,Wehner2018} and secure voting~\citep{Clarkson2008}. Classical protocols, such as Diffie-Hellman and RSA~\citep{Diffie1976,Rivest1978}, rely on computational hardness assumptions vulnerable to quantum algorithms~\citep{Shor1999} and side-channel attacks~\citep{Kocher1999}. These vulnerabilities highlight an urgent need for cryptographic frameworks that transcend classical limitations and offer robust, future-proof security.

Quantum cryptography emerges as a transformative solution, delivering information-theoretic security rooted in the fundamental laws of quantum mechanics. This security is based on principles such as the uncertainty relation~\citep{Bennett1992,Christandl2009,Portmann2022}, entanglement monogamy~\citep{Gisin2002,Masanes2009,Colbeck2011,Scarani2014,Pirandola2020,Portmann2022}, and the no-cloning theorem~\citep{bennett1984quantum,Gisin2002}. Unlike traditional quantum key distribution (QKD), which assumes trusted devices, device-independent QKD (DI-QKD) treats measurement devices as untrusted ``black boxes''. It certifies security through the violation of Bell inequalities~\citep{Bell1964,Toner2009,Einstein1935,Bell1987,Barrett2005a}, a hallmark of quantum nonlocality~\cite{Acin2007,Bell1964}. Bipartite DI-QKD~\citep{ekert1991quantum,Barrett2005b,Acin2007,Pironio2009,Barrett2012,Vazirani2014,miller2016robust,arnon2018practical,woodhead2021device,sekatski2021device} has achieved notable success. However, adapting it to multiparty settings, known as device-independent conference key agreement (DICKA)~\citep{Ribeiro2018,Holz2019,Ribeiro2019,Holz2020,Grasselli2023,Grasselli2021,Woodhead2018}, introduces formidable obstacles. Conventional multiparty approaches often require genuine multipartite entanglement (GME), such as Greenberger-Horne-Zeilinger (GHZ) states. These states are notoriously fragile, sensitive to noise, and experimentally demanding to produce and maintain over large networks~\cite{Ribeiro2018}.

Alternatively, DICKA can be realized by chaining bipartite DI-QKD links between a central node and each other party \citep{Epping2017}. In a recent advancement, Wooltorton et al. ~\citep{Wooltorton2025} propose a DICKA protocol utilizing biseparable states and a single Bell inequality, mirroring the structure of concatenated bipartite schemes~\citep{Carrara2021,Grasselli2018}. Despite meeting the definition of biseparability, the state may exhibit operational inconsistencies, as noted in~\citep{Navas2020}. Additionally, \citep{Das2021,Horodecki2022} demonstrates that multipartite private states-those enabling secret key extraction via local measurements-are inherently genuinely entangled. Consequently, biseparable states that are tensor-stable (i.e., remain biseparable regardless of the number of copies) are not suitable for conference key agreement. The biseparable state in~\citep{Wooltorton2025} is not tensor stable and hence have the potential to distill genuine entangled states from it. As a result, current state-of-the-art DICKA protocols cannot avoid relying on genuine multipartite entangled states. Many of these protocols depend on maximally entangled states \citep{Ribeiro2018,Holz2019,Ribeiro2019,Holz2020}, which demand significant experimental resources. In contrast, non-maximally entangled genuine states, which require fewer resources, present a more practically viable option.  Motivated by this observation, our protocol employs a genuine non-maximally entangled Hardy state and extracts the secret key from the measurement inputs rather than the measurement outcomes. This allows us to explore a complementary direction for device-independent conference key agreement, where the security certification arises from the multipartite Hardy paradox and the resulting protocol can exhibit robustness against imperfections in the random number generators used to select the measurement settings.

In this article, we introduce a novel DICKA protocol for $N$ parties that leverages non-maximally entangled genuine multipartite states. Our approach utilizes the multipartite Hardy paradox, a powerful alternative to Bell inequalities for certifying genuine multipartite nonlocality~\cite{Rahaman2014,Adhikary2024pla}. Originally formulated as a ladder of logical constraints that bipartite quantum systems~\cite{Hardy1993} paradoxically satisfy (in the context of local realistic theories), the Hardy paradox extends naturally to multiparty settings. It provides a robust framework for security certification without requiring inequality-based bounds. Building on the bipartite approach of~\citep{Rahaman2015}, which uses measurement settings for key generation, our protocol employs measurement settings to derive keys in a multiparty setting under maximal violation. This differs from traditional DI-QKD, which relies solely on measurement outcomes.

 The central motivation for extracting the key from settings rather than outcomes lies in their origin: settings are locally generated by the parties’ random number generators (RNGs), while outcomes come from systems distributed by the untrusted source. Even small imperfections in RNGs can endanger security~\cite{bouda2012weak,huber2013weak}, and the vulnerability is especially severe in the bipartite scenario, when the key is drawn from outcomes~\cite{Rahaman2015}. To benchmark this effect, we consider the tripartite Holz~\cite{Holz2020} and Parity-CHSH~\cite{Ribeiro2019} inequalities, identified by Grasselli et al. \cite{Grasselli2023} as yielding the optimal key rates for DICKA. Strikingly, for these inequalities the DICKA key rate vanishes once the RNG bias exceeds $0.09$, whereas our Hardy-based input-key protocol continues to sustain positive rates well beyond this regime.

Apart from reducing the protocol's dependence on maximal GME, our method achieves resilience against collective attacks through self-testing properties inherent in device-independent frameworks~\cite{mayers1998quantum,mayers2004self}. A particularly striking feature of our protocol is that, by exploiting the Hardy paradox, any two participants can distill a shared secret key at a substantially higher rate than that achievable by the entire $N$-party group.  This enhancement arises because the paradox enforces stronger, more resilient nonlocal correlations between each pair of parties. This unique capability, inherent to the multipartite Hardy paradox, allows for tailored key distribution within the group, where these individual keys can be combined using the XOR operation to achieve the full conference key agreement. By utilizing non-maximally entangled states and self-testing properties, we demonstrate positive key rates, advancing device-independent multiparty cryptography.

The subsequent sections of this paper are structured as follows. In section \ref{sec:genHardy}, we initially present multipartite Hardy's non-locality argument. In section \ref{sec:selftestN}, we briefly discuss the self-testing property of the quantum state that violates the paradox maximally. Moving on, in section \ref{sec:proto}, we introduce a protocol for conference key agreement among $N$-parties. Later, in section \ref{sec:key} we provide detail description of key rate and a possible approach of boosting it. In Section~\ref{apnA}, we further demonstrate resilience to realistic imperfections by establishing the robustness of our protocol against noise in the correlations through robust self-testing of both the state and the measurements.  In Section~\ref{security}, we analyze the performance of the protocol in the presence of noise and quantify the resulting secret key rate. Section~\ref{sec:biased} highlights that the distribution of measurement settings--particularly when influenced by biases in the random number generators--has a critical impact on the robustness of CKA . Lastly, we summarize our work in the concluding section \ref{sec:conclu}. 

\section{Multipartite Hardy's Paradox}
\label{sec:genHardy}
Consider $N$ distant parties $A_1$, $A_2$, $\ldots$ and $A_N$ sharing a multipartite quantum system. Each party $A_i$ has a choice of two dichotomic observables, $A_i^0$ and $A_i^1$, with binary outcomes $\{+1,-1\}$. We consider a generalization of Hardy's paradox for $N$ parties, as proposed in \cite{Rahaman2014}, defined by the following conditions:

\begin{eqnarray}\label{Hardy3}
        &p_N= P(+1,+1,\dots,+1|A_1^0,A_2^0,\dots,A_N^0), \label{eq:H3} \\
            &\begin{aligned} \label{tri}
			&\mbox{for~}i=1,2,\dots,N,\quad P(+1,+1|A_i^1,A_{i+1}^0)=0,\\
			&P(-1,-1,\dots,-1|A_1^1,A_2^1,\dots,A_N^1)=0,
		\end{aligned}  
            \end{eqnarray}

with the convention that $N+1\equiv 1$. This set of conditions cannot be satisfied by any fully local or no-signaling bi-local correlation~\citep{Bancal2013}, making it a genuine nonlocal correlation~\citep{Adhikary2024pla,Adhikary2025}. Furthermore, any genuine nonlocal correlation satisfying the conditions in \eqref{tri} also implies:
\begin{eqnarray}\label{keyeqp}
   &P(+1,+1,\dots,+1|A_1^{k_1},A_2^{k_2},\dots,A_N^{k_N})=0 \\&\mbox{ for } \{k_1,k_2,\dots ,k_N\} \neq\{0,0,\ldots,0\} \mbox{ or } \{1,1,\ldots,1\}\nonumber 
\end{eqnarray}	
	
\section{Self-testing of $N$-qubit Hardy state}\label{sec:selftestN}
Self-testing certifies quantum states and measurements solely from the observed correlations, without trusting the devices~\citep{supic2020selftesting}. Here we summarize the self-testing of the Hardy state achieving the maximal probability $p_N^{max}$, following~\cite{Adhikary2024}.

Consider the observables $A_i^0$ and $A_i^1$ for the party $A_i$. Without loss of generality, we set $A_i^0=\sigma_z$. The observable $A_i^1$ is defined as $\ket{A_{i,+}^1}\bra{A_{i,+}^1}-\ket{A_{i,-}^1}\bra{A_{i,-}^1}$, where:
\begin{eqnarray}
  \ket{A_{i,+}^1}&=&\cos \frac{\alpha_i}{2}\ket{0} + e^{\iota \phi_i}\sin \frac{\alpha_i}{2}\ket{1}\nonumber\\
  \ket{A_{i,-}^1}&=&-\sin \frac{\alpha_i}{2}\ket{0} + e^{\iota \phi_i}\cos \frac{\alpha_i}{2}\ket{1}\nonumber
\end{eqnarray}
with $\alpha_i \in[0, \pi)$ and $\phi_i \in [0, 2\pi)$.

The $N$-qubit Hardy state $\ket{\psi^H_N}$ is expressed as:
\begin{equation}
    \ket{\psi^H_N} = \sum_{j_1, \dots, j_N} c_{j_1, \dots, j_N} \ket{A_{1,j_1}^1} \otimes \ket{A_{2,j_2}^1} \otimes \dots \otimes \ket{A_{N,j_N}^1}
    \label{generalform}
\end{equation}
where $j_1,\dots,j_N \in \{+,-\}$.

For the violation of $p_N^{max}$, $\alpha_i=\alpha$ for all $i$, yielding $A_i^1 =A^1$. The state $\ket{\psi^H_N}$ then becomes:
\begin{eqnarray} \label{hardystateN}
        \ket{\psi^H_N} &=&\frac{1}{\sqrt{l_N}} \Big[\ket{A_{+}^1 \dots A_{+}^1} + \cot{\frac{\alpha}{2}} \ket{\mbox{Perm}(A_{-}^1A_{+}^1\dots A_{+}^1)} \nonumber \\
        &+& \cot^2{\frac{\alpha}{2}} \ket{\mbox{Perm}(A_{-}^1A_{-}^1\dots A_{+}^1)} \nonumber \\
        &+& \dots + \cot^{N-1}{\frac{\alpha}{2}} \ket{\mbox{Perm}(A_{-}^1\dots A_{-}^1 A_{+}^1)} \Big] \nonumber
\end{eqnarray}
where $l_N=\sum_{k=0}^{N-1}\binom Nk
\cot^{2k}\!\tfrac\alpha2$, and $\text{Perm}$ denotes permutations of the projectors. Here, $\alpha=2 \arccos \sqrt{t_r}$, where $t_r$ is the positive root of the polynomial $x^{N+1}-2x+1$ other than $1$ (there is only one positive root of the equation other than $1$). The probability $p_N^{max}$ is $\frac{t_r^N (1 - t_r)^N}{1 - t_r^N}$.

Moreover, any state $\rho$ achieving \(p_N^{\max}\) admits local isometries \(\{\mathcal O_i\}\)~\cite{Adhikary2024} such that
\begin{align*}
 (\bigotimes_{i=1}^N\mathcal O_i)\,
\bigl(\rho\otimes\varrho_{1'\dots N'}\bigr)\,
(\bigotimes_{i=1}^N\mathcal O_i^\dagger)
=\lvert\psi^H_N\rangle\langle\psi^H_N\rvert
\otimes
\varrho'_{1'\dots N'},   
\end{align*}
thus self-testing \(\lvert\psi^H_N\rangle\).

In the special case $N=3$, one obtains:
   \begin{eqnarray} \label{hardystate3}
        \ket{\psi^H_3} &=&\frac{1}{\sqrt{l_3}} \Big[\ket{A_{+}^1A_{+}^1 A_{+}^1} \nonumber \\&+& \cot{\frac{\alpha}{2}} (\ket{A_{-}^1A_{+}^1A_{+}^1} + \ket{A_{+}^1A_{-}^1A_{+}^1} +\ket{A_{+}^1A_{+}^1A_{-}^1}) \nonumber \\
        &+& \cot^2{\frac{\alpha}{2}} (\ket{A_{-}^1A_{-}^1A_{+}^1} + \ket{A_{-}^1A_{-}^1A_{+}^1} +\ket{A_{+}^1A_{-}^1A_{-}^1}) \Big] \nonumber
\end{eqnarray}
with $l_3=1+2\cot^2{\frac{\alpha}{2}}+2\cot^4{\frac{\alpha}{2}}$ and $\alpha=2 \arccos \sqrt{t_r}$. Here, $p_3^{max}=\dfrac{t_{r}^3(1-t_{r})^3}{1-t_{r}^3}$ where $t_r=\tfrac{1}{3} \left(\small{\sqrt[3]{17+3 \sqrt{33}}}-\tfrac{2}{\sqrt[3]{17+3 \sqrt{33}}}-1\right)$.

\section{Conference Key Agreement}\label{sec:proto}
We consider $N$ distant parties, $A_1, A_2, \dots, A_N$, aiming to generate a secure conference key using public classical communication channels. The device-independent conference key agreement (DICKA) protocol proceeds as follows:

\begin{itemize}
    \item \textbf{Step 1 (Initial Phase):} Each party receives one qubit from a shared $N$-qubit entangled state per round.
    \item \textbf{Step 2 (Measurement Phase):} Each party $A_i$ randomly selects a measurement basis, either $A_i^0$ or $A_i^1$, measures their qubit, and records the outcome.
    \item \textbf{Step 3 (Eavesdropping Check):} For a random subset of runs, parties announce their measurement bases and outcomes to verify the Hardy state $\ket{\psi^H_N}$ via its maximal violation $p_N^{\text{max}}$. If achieved, the protocol proceeds; otherwise, it aborts.  In practice, the maximal violation of the Hardy paradox cannot be achieved exactly due to statistical fluctuations and experimental imperfections. Therefore, the protocol does not require the ideal Hardy conditions to be satisfied exactly. Instead, the parties estimate the relevant probabilities and verify that they lie within admissible tolerances $(\varepsilon_1,\varepsilon_2)$, as discussed in Sec.\ref{security}. The protocol proceeds whenever the observed correlations satisfy these noisy Hardy constraints, while if the deviations exceed these tolerances the protocol aborts .
    \item \textbf{Step 4 (Key Generation):} For remaining runs, parties announce only outcomes. Runs where all obtain $+1$ are selected.   Now $A_1$ selects a random subset of the runs for which measurement setting is $A_1^1$. The fraction of runs $A_1$ keeps is
\begin{equation*}
\frac{P(+1,\ldots,+1|A_1^0,\ldots,A_N^0)}{P(+1,\ldots,+1|A_1^1,\ldots,A_N^1)}.    
\end{equation*}
$A_1$ then publicly communicates the list of these selected runs to $A_2$, $\dots$, $A_N$. As $P(+1,\ldots,+1|A_1^0,\ldots,A_N^0) < P(+1,\ldots,+1|A_1^1,\ldots,A_N^1)$ for $t\in(0,1)$, $A_1$ keeps $\frac{P(+1,\ldots,+1|A_1^0,\ldots,A_N^0)}{P(+1,\ldots,+1|A_1^1,\ldots,A_N^1)}$ many rounds to make the guessing probability for Eve exactly $\frac{1}{2}$.

    \item \textbf{Step 5 (Key Assignment):} For selected runs, parties assign bit values: 0 for $A_i^0$, 1 for $A_i^1$, yielding identical keys in the ideal case.
\end{itemize}

The device-independent framework quantifies all eavesdropper interventions, including device manipulations. The multipartite Hardy test~\eqref{Hardy3}, with maximal violation $p_N^{\text{max}}$, ensures a unique quantum probability distribution~\citep{Adhikary2024}.

In the $N$-partite scenario we have,
\begin{equation}
    \begin{aligned}
        0 &< P(+1, \dots, +1 | A_1^0, \dots, A_N^0) = \frac{t_r^N (1 - t_r)^N}{1 - t_r^N}, \\
        0 &< P(+1, \dots, +1 | A_1^1, \dots, A_N^1) = \frac{(1 - t_r)^N}{1 - t_r^N},
    \end{aligned}
    \label{keyeq}
\end{equation}
where $t_r$ is the positive root of $x^{N+1} - 2x + 1$ other than 1 \cite{Adhikary2024}. These conditions ensure key generation only when all parties select the same basis, securing the protocol against collective memory attacks.

  The above protocol is valid only under the ideal condition \eqref{tri}. In a realistic setting, however, this condition is inevitably violated. To establish the correctness of the protocol in such cases, one must provide a robust self-test of $\ket{\psi^H_N}$, as discussed in Section~\ref{apnA}.

\section{Key rate calculations}\label{sec:key}

Considering the probability of obtaining the non-dropped outcomes, we have:
\begin{equation*}
  P_\text{nd}(+1,\dots,+1|A_1^0,\dots,A_N^0) + P_\text{nd}(+1,\dots,+1|A_1^1,\dots,A_N^1)= \dfrac{2t_{r}^N(1-t_{r})^N}{1-t_{r}^N}.  
\end{equation*}

To determine the fraction of total runs that contribute to the key, this result is multiplied by the probability of choosing each measurement setting, which is:
\begin{equation*}
P(A_1^0,\dots,A_N^0) = P(A_1^1,\dots,A_N^1) = \frac{1}{2^N}.
\end{equation*}
Thus, the overall key rate is: $\dfrac{t_{r}^N(1-t_{r})^N}{2^{N-1}(1-t_{r}^N)}$. For $N=3$ the key rate is $0.004548$. In Figure \ref{fig:keyrate}, we provide the plot of the key rate respect to number of party $N$ in $\log_{10}$ scale. Note that the key rate becomes very small for large number of parties. We name this scheme as ``\emph{Protocol 1}''.

\begin{figure}[htbp]
\centering
\begin{tikzpicture}
\begin{axis}[
    ymode=log,
    log basis y=10,
    xlabel={\textbf{Number of parties $N$}},
    ylabel={\textbf{Key rate}},
    width=8.5cm,
    height=6.5cm,
    xmin=3, xmax=10,
    xtick={3,4,5,6,7,8,9,10},
    ymin=1e-10, ymax=1e-2,
    ytickten={-10,-9,...,-2},
    minor tick num=5,
    minor tick num=2,
    major grid style={gray!20},
    minor grid style={gray!10},
    mark size=3pt,
    font=\footnotesize,
    scaled ticks=false,
    yticklabel style={/pgf/number format/fixed}
]

\addplot+[
    color=blue!70!black,
    mark=*,
    thick,
    mark options={fill=blue!70!black},
    smooth
]
coordinates {
  (3, 0.00454846)
  (4, 0.000523446)
  (5, 0.0000630921)
  (6, 7.75355e-6)
  (7, 9.6129e-7)
  (8, 1.19681e-7)
  (9, 1.49305e-8)
  (10, 1.86447e-9)
};


\end{axis}
\end{tikzpicture}
\caption{Key rate plot for \emph{Protocol 1}, demonstrating the protocol's performance when all participants are involved in the key generation round.}
\label{fig:keyrate}
\end{figure}


Now we will provide a scheme under which we will get better key rate.   This scheme, referred to as ``\emph{Protocol 2}'', exploits the fact that for $p^{\max}_N$ the Hardy state $\ket{\psi^H_N}$ is symmetric and the local observables satisfy $A^k_i=A^k_j$ for all $i\neq j$~\cite{Adhikary2024}. Introducing the swap operator $\mathrm{S}_{A_i,A_j}$, defined by $\ket{\phi}_{A_i}\otimes\ket{\xi}_{A_j}\mapsto \ket{\xi}_{A_i}\otimes\ket{\phi}_{A_j}$, we note that $\ket{\psi^H_N}$ is invariant under $\mathrm{S}_{A_i,A_j}$ for any pair $i,j$. It follows that
\begin{equation}
\begin{split}
  &P(+1,+1|A_1^1,A_2^0)\\&=\tr\!\left(\proj{\psi^H_N}(A_{1,+}^1\otimes A_{2,+}^0)\right)\\
&=\tr\!\left(\mathrm{S}_{A_2,A_j}\proj{\psi^H_N}\mathrm{S}_{A_2,A_j}(A_{1,+}^1\otimes A_{2,+}^0)\right)\\
&=\tr\!\left(\proj{\psi^H_N}\mathrm{S}_{A_2,A_j}(A_{1,+}^1\otimes A_{2,+}^0)\mathrm{S}_{A_2,A_j}\right)\\
&=\tr\!\left(\proj{\psi^H_N}(A_{1,+}^1\otimes A_{j,+}^0)\right)\\
&=P(+1,+1|A_1^1,A_{j}^0),   
\end{split}
\end{equation}so that, in general, $P_{A_iA_j}(+1,+1|A_i^1,A_j^0)=0,\forall i\neq j.$

Hence we have the following condition,

\begin{equation}\label{newcon}
\begin{split}
P_{A_iA_j}(+1,+1|A_i^0,A_j^0)&=\dfrac{t_{r}^N(1-t_{r})^{N-1}}{1-t_{r}^N} \\ 
P_{A_iA_j}(+1,+1|A_i^1,A_j^1)&=\dfrac{(1-t_{r})^{N-1}}{1-t_{r}^N} \\
P_{A_iA_j}(+1,+1|A_i^1,A_j^0)&=0. 
\end{split}
\end{equation}

Clearly from \eqref{newcon} it is apparent that any two party $A_i$ and $A_j$ can establish a key. As $P_{A_iA_j}(+1,+1|A_i^0,A_j^0) < P_{A_iA_j}(+1,+1|A_i^1,A_j^1)$ we will use the dropping strategy. Hence the key rate will be,
\begin{eqnarray}
   &\frac{1}{4}(P_{A_iA_j}(+1,+1|A_i^0,A_j^0) + P_{A_iA_j}(+1,+1|A_i^1,A_j^1))\nonumber\\ &= \dfrac{t_{r}^N(1-t_{r})^{N-1}}{2(1-t_{r}^N)}\nonumber 
\end{eqnarray}

Suppose $\mathsf{K}_{A_1A_2}$ is the key between $A_1$ and $A_2$. Similarly $A_1$ can have key $\mathsf{K}_{A_1A_3}$, $\dots$, $\mathsf{K}_{A_1A_N}$ with $A_3$, $\dots$, $A_N$ respectively. Now $A_1$ can publicly announce $\mathsf{K}_{A_1A_2}\oplus \mathsf{K}_{A_1A_3}$ and $A_3$ can know $\mathsf{K}_{A_1A_2}$ simply by $\mathsf{K}_{A_1A_2}\oplus \mathsf{K}_{A_1A_3}\oplus \mathsf{K}_{A_1A_3}$. Note that Eve has no information about $\mathsf{K}_{A_1A_2}$. Hence we have a key between 
$A_1$, $A_2$ and $A_3$. $A_1$ can do the same thing with $A_4$, $\dots$, $A_N$ respectively. Hence we have a key $\mathsf{K}_{A_1A_2}$ between $A_1$, $A_2$, $\dots$, $A_N$. Note that in this protocol the key rate is, $\dfrac{t_{r}^N(1-t_{r})^{N-1}}{2(1-t_{r}^N)}$ which is larger than the previous key rate $\dfrac{t_{r}^N(1-t_{r})^N}{2^{N-1}(1-t_{r}^N)}$. For $N=3$ the key rate in \emph{Protocol 2} is $0.02$ which is better than $0.004548$ of \emph{Protocol 1}. In Figure \ref{fig:betterkeyrate}, we provide the plot of the key rate in \emph{Protocol 2} respect to number of party $N$.  This protocol illustrates how the symmetry of the Hardy state under maximal violation enables the extraction of identical pairwise correlations from a single multipartite resource, allowing higher pairwise key rates in the ideal scenario .

\begin{figure}[htbp]
\centering
\begin{tikzpicture}
\begin{axis}[
    ymode=log,
    log basis y=10,
    xlabel={\textbf{Number of parties $N$}},
    ylabel={\textbf{Key rate}},
    width=8.5cm,
    height=6.5cm,
    xmin=3, xmax=10,
    xtick={3,4,5,6,7,8,9,10},
    ymin=1e-7, ymax=1e-1,
    ytickten={-7,-6,...,-2},
    minor tick num=5,
    minor tick num=2,
    major grid style={gray!20},
    minor grid style={gray!10},
    mark size=2.5pt,
    font=\footnotesize,
    scaled ticks=false,
    yticklabel style={/pgf/number format/fixed}
]

\addplot+[
    color=blue!70!black,
    mark=*,
    thick,
    mark options={fill=blue!70!black},
    smooth
]
coordinates {
  (3, 0.0199358)
  (4, 0.00435108)
  (5, 0.00102727)
  (6, 0.000250184)
  (7, 0.0000617718)
  (8, 0.0000153496)
  (9, 3.82597e-6)
  (10, 9.55078e-7)
};

\end{axis}
\end{tikzpicture}
\caption{A comparison of key rates between \emph{Protocol 1} (\ref{fig:keyrate}) and \emph{Protocol 2}, showcasing the superior efficiency of \emph{Protocol 2}. By generating pairwise keys between a central party $A_1$ and each $A_i$ ($i=2,\dots,N$), and then using XOR operations to establish a shared conference key, \emph{Protocol 2} achieves higher key rates than \emph{Protocol 1}.}
\label{fig:betterkeyrate}
\end{figure}

\section{Robust self-testing of Hardy state and measurement} \label{apnA}

The $N$-partite Hardy paradox~\ref{Hardy3} features the zero-probability constraints~\ref{tri}, which also underpin our cryptographic protocol. Of course, demanding strictly vanishing probabilities is unrealistic in any physical experiment. It is therefore essential to assess how robust a given self-testing statement is against small imperfections. Let us recall the SWAP method \cite{yang2014robust,bancal2015physical,Chen2023quantumcorrelations} and explain how to use it to obtain numerical robustness bounds for the Hardy state and the associated measurements.

We consider local operators $\mathsf{S}_{A_iA'_i}$ acting jointly on a black-box system $A_i$ and a trusted auxiliary system $A'_i$, and define
\begin{equation}
    \mathcal{S}\!\left(\rho_{A_1\ldots A_N}\otimes\proj{0\ldots 0}_{A'_1\ldots A'_N}\right)\mathcal{S}^\dagger,
\end{equation}
where $\mathcal{S}=\mathsf{S}_{A_1A'_1}\otimes\cdots\otimes\mathsf{S}_{A_NA'_N}$. In the ideal case where the actual observables $A_i^j$ coincide with the reference ones $\widetilde{A}_i^j$, we choose $\mathsf{S}_{A_iA'_i}$ to swap the Hilbert spaces $\mathcal{H}_{A_i}$ and $\mathcal{H}_{A'_i}$. The fidelity
\begin{equation}\label{eq:statefidelity}
    F=\bra{\widetilde{\psi}}\rho_{\text{\tiny SWAP}}\ket{\widetilde{\psi}},
\end{equation}
between the reference state $\ket{\widetilde{\psi}}$ and the \emph{swapped} state
\begin{equation}\label{eq:rho_swap}
    \rho_{\text{\tiny SWAP}}=
    \tr_{A_1\ldots A_N}\!\left[
    \mathcal{S}\left(\rho_{A_1\ldots A_N}\otimes\proj{0\ldots 0}_{A'_1\ldots A'_N}\right)\mathcal{S}^\dagger
    \right],
\end{equation}
then quantifies how close the unknown shared state $\rho_{A_1\ldots A_N}$ is to $\ket{\widetilde{\psi}}$. Note that, the entries of $\rho_{\text{\tiny SWAP}}$ are given by linear combinations of correlation terms from the set $d=\{\tr(\mathbb{1}\rho_{A_1\ldots A_N}),\tr(A^0_1\rho_{A_1\ldots A_N}),\ldots, \tr(A^0_1A^1_2\dots A^0_N\rho_{A_1\ldots A_N}),\ldots\}$. The fidelity $F$ is hence a linear combination of these moments. 

As this problem is numerically challenging for general $\ket{\psi^H_N}$, we will provide the result only for $\ket{\psi^H_3}$.

\subsection{Robust self-test of $\ket{\psi^H_3}$ and the associated measurement}\label{robust3}
Consider the tripartite ($N=3$) Hardy paradox~\ref{Hardy3}. Now, the maximal Hardy success is
\[
p^{\max}_3=\frac{t_r^{\,3}(1-t_r)^{3}}{1-t_r^{\,3}},\
t_r=\frac{1}{3}\left(\sqrt[3]{17+3\sqrt{33}}-\frac{2}{\sqrt[3]{17+3\sqrt{33}}}-1\right).
\]
At this point, the Hardy state can be written as
\begin{align*}
 \ket{\psi^H_3}
= c_0\ket{000}+c_1(\ket{001}+\ket{010}+\ket{100})
+ c_2(\ket{011}+\ket{110}+\ket{101})+c_3\ket{111},   
\end{align*}

with
\begin{align*}
c_0=\frac{a^3 b^3}{\sqrt{1-a^6}},
c_1=-\frac{a^4 b^2}{\sqrt{1-a^6}},
c_2=\frac{a^5 b}{\sqrt{1-a^6}},
c_3=\sqrt{1-a^6},
\end{align*}
where
\[
a=\frac{\sqrt{-2-\sqrt[3]{17+3\sqrt{33}}+\left(17+3\sqrt{33}\right)^{2/3}}}{\sqrt{3}\,\sqrt[6]{17+3\sqrt{33}}},\quad
b=\sqrt{1-a^2}.
\]

We will now construct the local isometry that will work as swap between the Hilbert space $A_i$ and auxiliary Hilbert space $A'_i$. We will follow the same procedure as depicted in \cite{yang2014robust,bancal2015physical,Chen2023quantumcorrelations}. For each $A_i$ we define
\begin{align}\label{eq:localiso}
    \mathsf{S}_{A_iA'_i}:=U_{A_iA'_i}V_{A_iA'_i}U_{A_iA'_i},
\end{align}
where
\begin{align}
   U_{A_iA'_i}&:=\mathbb{1}_{A_i}\otimes\proj{0}_{A'_i}+X_{A_i}\otimes\proj{1}_{A'_i},\nonumber\\
   V_{A_iA'_i}&:=\frac{\mathbb{1}_{A_i}+Z_{A_i}}{2}\otimes\mathbb{1}_{A'_i}
   +\frac{\mathbb{1}_{A_i}-Z_{A_i}}{2}\otimes\sigma_{x,{A'_i}}.\nonumber
\end{align}

Now consider, $\rho_{A_1A_2A_3}$ to be a general tripartite qudit state. Replacing \eqref{eq:localiso} in \eqref{eq:rho_swap}, we get

\begin{equation}\label{eq:rho_swap3}
\begin{split}
     \rho_{\text{\tiny SWAP}}&=
    \tr_{A_1A_2A_3}\!\left[
    \mathcal{S}\left(\rho_{A_1A_2A_3}\otimes\proj{000}_{A'_1A_2A'_3}\right)\mathcal{S}^\dagger
    \right]\\
    &=\sum_{\mathsf{ijklst}}\mathsf{C}_{\mathsf{ijklst}}\;
\ket{\mathsf{i}}\bra{\mathsf{l}}\otimes
\ket{\mathsf{j}}\bra{\mathsf{s}}\otimes
\ket{\mathsf{k}}\bra{\mathsf{t}},
\end{split}
\end{equation}

where
\[
\mathsf{C}_{\mathsf{ijklst}}=\frac{1}{64}\,
\tr\!\Big[\big(\mathsf{M}^{A_1}_{\mathsf{il}}\otimes \mathsf{M}^{A_2}_{\mathsf{js}}\otimes \mathsf{M}^{A_3}_{\mathsf{kt}}\big)\rho_{A_1A_2A_3}\Big],
\]
and
\begin{align*}
\mathsf{M}^{A_1}_{\mathsf{il}}&=(\mathbb{1}+Z_{A_1})^{1-\mathsf{l}}(X_{A_1}-Z_{A_1}X_{A_1})^{\mathsf{l}}
(\mathbb{1}+Z_{A_1})^{1-\mathsf{i}}(X_{A_1}-X_{A_1}Z_{A_1})^{\mathsf{i}},\\
\mathsf{M}^{A_2}_{\mathsf{js}}&=(\mathbb{1}+Z_{A_2})^{1-\mathsf{s}}(X_{A_2}-Z_{A_2}X_{A_2})^{\mathsf{s}}
(\mathbb{1}+Z_{A_2})^{1-\mathsf{j}}(X_{A_2}-X_{A_2}Z_{A_2})^{\mathsf{j}},\\
\mathsf{M}^{A_3}_{\mathsf{kt}}&=(\mathbb{1}+Z_{A_3})^{1-\mathsf{t}}(X_{A_3}-Z_{A_3}X_{A_3})^{\mathsf{t}}
(\mathbb{1}+Z_{A_3})^{1-\mathsf{k}}(X_{A_3}-X_{A_3}Z_{A_3})^{\mathsf{k}}.
\end{align*}
Substituting \eqref{eq:rho_swap3} into ~\eqref{eq:statefidelity} yields
\begin{equation}\label{figmeritstate}
\begin{split}
F =\;&
c_0\!\left(c_0 \mathsf{C}_{000000} + c_1 \mathsf{C}_{000001} + c_1 \mathsf{C}_{000010} + c_2 \mathsf{C}_{000011}
+ c_1 \mathsf{C}_{000100} + c_2 \mathsf{C}_{000101} + c_2 \mathsf{C}_{000110} + c_3 \mathsf{C}_{000111}\right) \\
&+ c_1\!\left(c_0 \mathsf{C}_{001000} + c_1 \mathsf{C}_{001001} + c_1 \mathsf{C}_{001010} + c_2 \mathsf{C}_{001011}
+ c_1 \mathsf{C}_{001100} + c_2 \mathsf{C}_{001101} + c_2 \mathsf{C}_{001110} + c_3 \mathsf{C}_{001111}\right) \\
&+ c_1\!\left(c_0 \mathsf{C}_{010000} + c_1 \mathsf{C}_{010001} + c_1 \mathsf{C}_{010010} + c_2 \mathsf{C}_{010011}
+ c_1 \mathsf{C}_{010100} + c_2 \mathsf{C}_{010101} + c_2 \mathsf{C}_{010110} + c_3 \mathsf{C}_{010111}\right) \\
&+ c_2\!\left(c_0 \mathsf{C}_{011000} + c_1 \mathsf{C}_{011001} + c_1 \mathsf{C}_{011010} + c_2 \mathsf{C}_{011011}
+ c_1 \mathsf{C}_{011100} + c_2 \mathsf{C}_{011101} + c_2 \mathsf{C}_{011110} + c_3 \mathsf{C}_{011111}\right) \\
&+ c_1\!\left(c_0 \mathsf{C}_{100000} + c_1 \mathsf{C}_{100001} + c_1 \mathsf{C}_{100010} + c_2 \mathsf{C}_{100011}
+ c_1 \mathsf{C}_{100100} + c_2 \mathsf{C}_{100101} + c_2 \mathsf{C}_{100110} + c_3 \mathsf{C}_{100111}\right) \\
&+ c_2\!\left(c_0 \mathsf{C}_{101000} + c_1 \mathsf{C}_{101001} + c_1 \mathsf{C}_{101010} + c_2 \mathsf{C}_{101011}
+ c_1 \mathsf{C}_{101100} + c_2 \mathsf{C}_{101101} + c_2 \mathsf{C}_{101110} + c_3 \mathsf{C}_{101111}\right) \\
&+ c_2\!\left(c_0 \mathsf{C}_{110000} + c_1 \mathsf{C}_{110001} + c_1 \mathsf{C}_{110010} + c_2 \mathsf{C}_{110011}
+ c_1 \mathsf{C}_{110100} + c_2 \mathsf{C}_{110101} + c_2 \mathsf{C}_{110110} + c_3 \mathsf{C}_{110111}\right) \\
&+ c_3\!\left(c_0 \mathsf{C}_{111000} + c_1 \mathsf{C}_{111001} + c_1 \mathsf{C}_{111010} + c_2 \mathsf{C}_{111011}
+ c_1 \mathsf{C}_{111100} + c_2 \mathsf{C}_{111101} + c_2 \mathsf{C}_{111110} + c_3 \mathsf{C}_{111111}\right).
\end{split}
\end{equation}

Now to assess the robustness of the self-testing statements relative to the reference state, we compute the worst-case fidelity by optimizing over all quantum realizations consistent with the observed value of the Hardy paradox.  In particular, we employ a relaxation method known as the NPA hierarchy \cite{Navascues2007,Navascues2008}, which provides an infinite sequence of outer approximations of the quantum set $
\mathcal{Q}_1 \supset \mathcal{Q}_2 \supset \dots \supset \mathcal{Q}_\ell \supset \dots$. Each level in this hierarchy is defined via SDP. It has been proven that these sets converge to the quantum set in the limit $ \ell \to \infty $, i.e., $\lim_{\ell\to\infty}\mathcal{Q}_\ell = \mathcal{Q}$ \cite{Navascues2007,Navascues2008}. 

\begin{figure*}[htbp]
    \centering
    \begin{subfigure}[t]{0.48\linewidth}
        \centering
        \includegraphics[width=\linewidth]{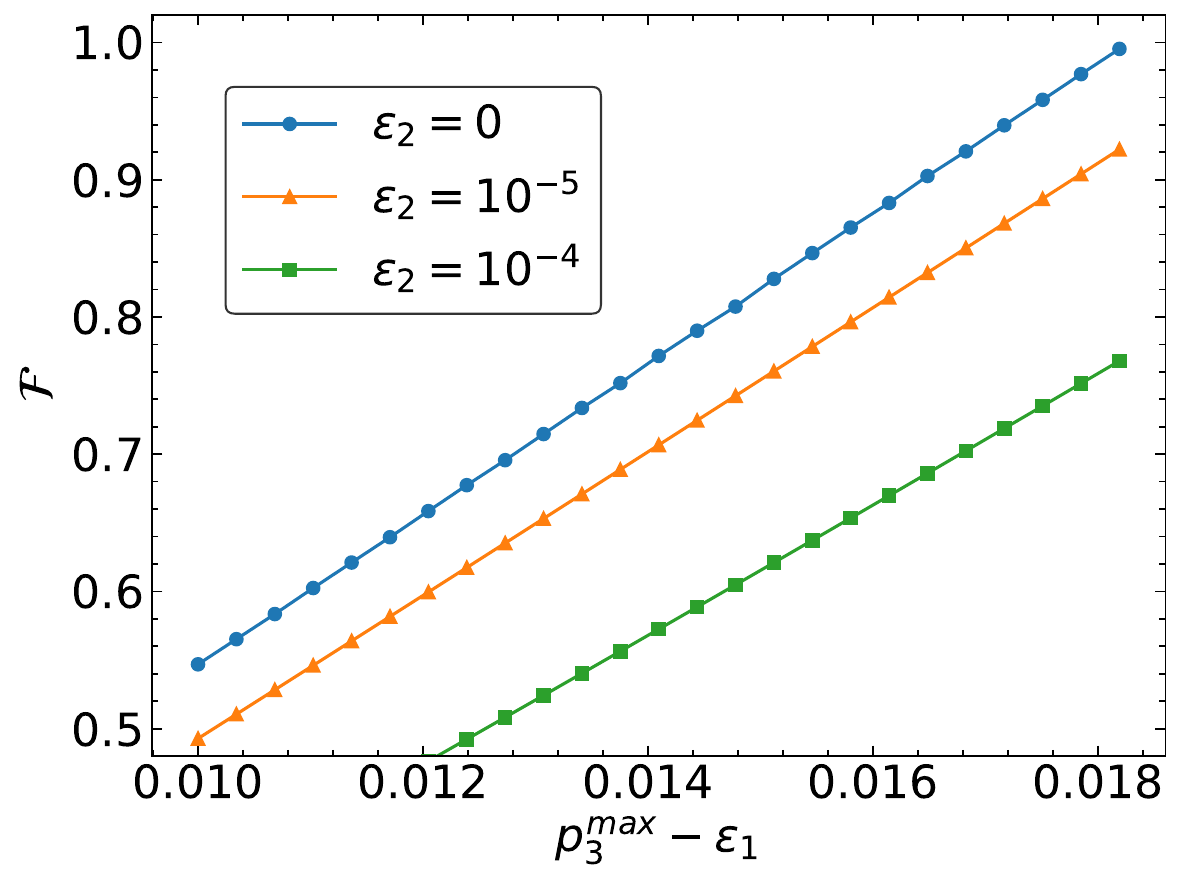}
        \caption{$\mathcal{F}$ corresponding to different values of $p^{max}_3 - \varepsilon_1$ and $\varepsilon_2$.}
        \label{fig:fidelity_hardy}
    \end{subfigure}\hfill
    \begin{subfigure}[t]{0.48\linewidth}
        \centering
        \includegraphics[width=\linewidth]{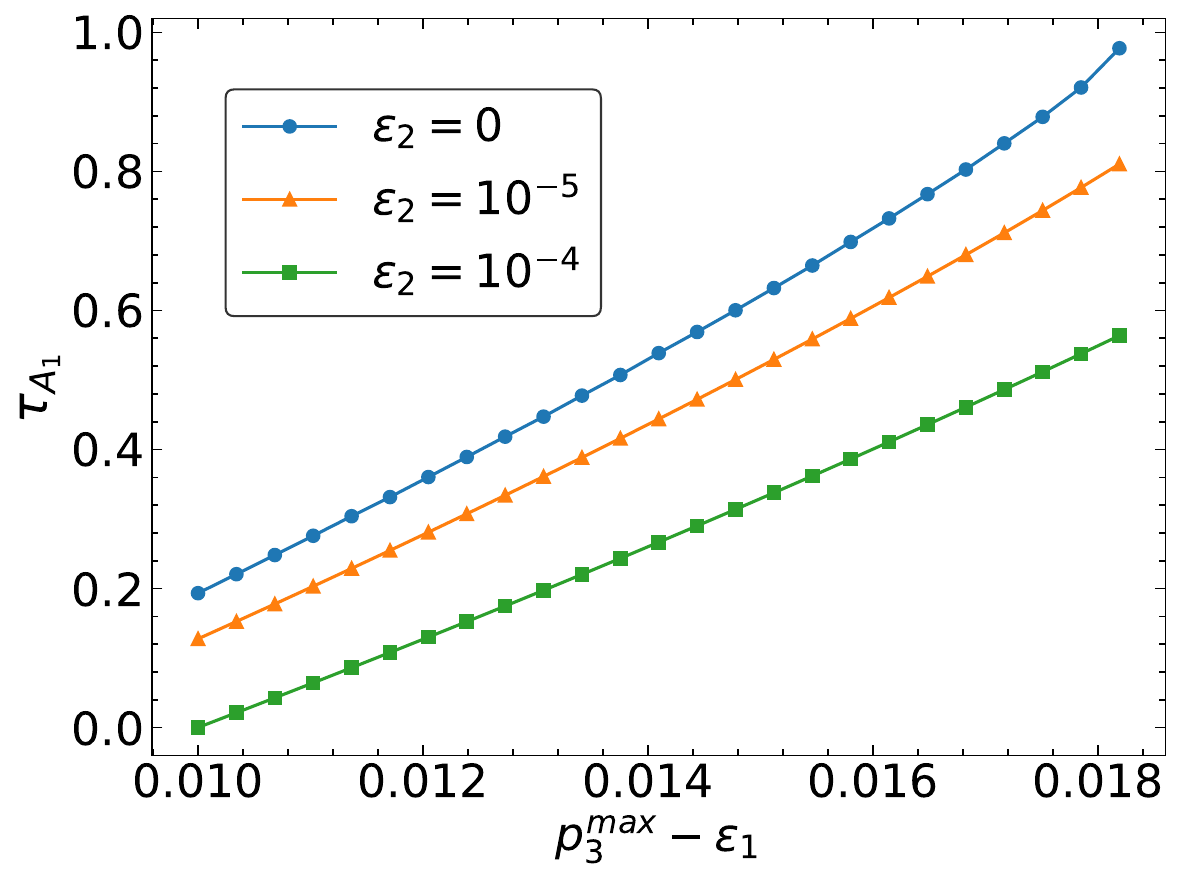}
        \caption{$\tau_{A_1}$ corresponding to different values of $p^{max}_3 - \varepsilon_1$ and $\varepsilon_2$.}
        \label{fig:fidelity_measure}
    \end{subfigure}
    \caption{  Plots illustrate the robustness of the self-testing result corresponding to the correlation that maximally violates the tripartite Hardy paradox. To quantify the quality of self-testing of the state, we adopt the fidelity with respect to the chosen reference state \eqref{figmeritstate} as the figure of merit. For the self-testing of the measurements, the relevant figure of merit is defined in \eqref{figmeritmeasure}. Because the reference measurements of $A_1$, $A_2$, and $A_3$ are similar, it is sufficient to display the plot for $A_1$ only. Throughout our analysis, the parameter $\varepsilon_2$ denotes the admissible deviation from the ideal zero-probability condition. All numerical results reported here are obtained using the level-6 outer approximation of the quantum set $\mathcal{Q}$.}
    \label{fig:robustness}
\end{figure*}

Now we solve the following optimization problem
\begin{equation}\label{eq:robust_state}
\begin{split}
     &\mathcal{F}=\min_{d\in \mathcal{Q}_\ell} F\\
     \text{s.t.}\;&\tr\left(\frac{\mathbb{1}+Z_{A_1}}{2} \otimes \frac{\mathbb{1}+Z_{A_2}}{2}\otimes \frac{\mathbb{1}+Z_{A_3}}{2} \rho_{A_1A_2A_3}\right)=p^{\max}_3-\varepsilon_1,\\
      &\tr\left(\frac{\mathbb{1}+X_{A_1}}{2} \otimes \frac{\mathbb{1}+Z_{A_2}}{2} \rho_{A_1A_2A_3}\right)\le\varepsilon_2,\\
      &\tr\left(\frac{\mathbb{1}+X_{A_2}}{2}\otimes \frac{\mathbb{1}+Z_{A_3}}{2} \rho_{A_1A_2A_3}\right)\le\varepsilon_2,\\
      &\tr\left(\frac{\mathbb{1}+Z_{A_1}}{2} \otimes \frac{\mathbb{1}+X_{A_3}}{2} \rho_{A_1A_2A_3}\right)\le\varepsilon_2,\\
      &\tr\left(\frac{\mathbb{1}-X_{A_1}}{2} \otimes \frac{\mathbb{1}-X_{A_2}}{2}\otimes \frac{\mathbb{1}-X_{A_3}}{2} \rho_{A_1A_2A_3}\right)\le\varepsilon_2,
\end{split}
\end{equation}
where $\varepsilon_1,\varepsilon_2\ge 0$ capture admissible deviations. The above optimization problem has been implemented in PYTHON using CVXPY, by setting the hierarchy level as $\ell=5$ \cite{gitmishra}. The resulting fidelity estimates are presented in Fig.~\ref{fig:fidelity_hardy}.

Now for the robust self-testing of measurements, we use the same figure of merit as considered in \cite{yang2014robust,bancal2015physical,Chen2023quantumcorrelations}

\begin{equation}\label{figmeritmeasure}
\mathsf{T}_{A_i}
=\frac{1}{2}[
P(0|A^0_i,\ket{0})+P(1|A^0_i,\ket{1})
+P(0|A^1_i,\ket{+})+P(1|A^1_i,\ket{-})
]-1,
\end{equation}
where
\begin{align}\label{Eq:P_A}
P(a|A^j_i,\ket{\phi})=
\tr\left\{
\frac{\mathbb{1}+(-1)^a A^j_i}{2}\otimes\mathbb{1}_{A'_i}\;
\big[\Phi_{A_iA'_i}\big(\tr_{\{A_1,A_2,A_3\}\setminus\{A_i\}}\rho_{A_1A_2A_3}\otimes\proj{\phi}_{A'_i}\big)\Phi_{A_iA'_i}^\dagger\big]
\right\}.
\end{align}
When the devices implement the reference measurements, each $\mathsf{T}_{A_i}$ equals $1$. As in our case the system $A_1$, $A_2$ and $A_3$ have same measurement, we only consider the analysis for system $A_1$. In our explicit case,
\begin{equation}
\begin{split}
P(0|A^0_1,\ket{0})
&=\tr\!\left[
\Big(\tfrac{\mathbb{1}+Z_{A_1}}{2}
+\tfrac{\mathbb{1}-Z_{A_1}}{2}\,X_{A_1}\,\tfrac{\mathbb{1}+Z_{A_1}}{2}\,X_{A_1}\,\tfrac{\mathbb{1}-Z_{A_1}}{2}\Big)\rho_{A_1A_2A_3}
\right],\\
P(1|A^0_1,\ket{1})
&=\tr\!\left[
\Big(X_{A_1}\,\tfrac{\mathbb{1}-Z_{A_1}}{2}\,X_{A_1}
+X_{A_1}\,\tfrac{\mathbb{1}+Z_{A_1}}{2}\,X_{A_1}\,\tfrac{\mathbb{1}-Z_{A_1}}{2}\,X_{A_1}\,\tfrac{\mathbb{1}+Z_{A_1}}{2}\,X_{A_1}\Big)\rho_{A_1A_2A_3}
\right],\\
P(0|A^1_1,\ket{+})
&=\tr\!\left[
\Big(\tfrac{\mathbb{1}+X_{A_1}}{2}
+\tfrac{\mathbb{1}-X_{A_1}}{2}\,Z_{A_1}\,\tfrac{\mathbb{1}+X_{A_1}}{2}\,Z_{A_1}\,\tfrac{\mathbb{1}-X_{A_1}}{2}\Big)\rho_{A_1A_2A_3}
\right],\\
P(1|A^1_1,\ket{-})
&=\tr\!\left[
\Big(Z_{A_1}\,\tfrac{\mathbb{1}+X_{A_1}}{2}\,Z_{A_1}
+Z_{A_1}\,\tfrac{\mathbb{1}-X_{A_1}}{2}\,Z_{A_1}\,\tfrac{\mathbb{1}+X_{A_1}}{2}\,Z_{A_1}\,\tfrac{\mathbb{1}-X_{A_1}}{2}\,Z_{A_1}\Big)\rho_{A_1A_2A_3}
\right].
\end{split}
\end{equation}
We then estimate the worst-case figures of merit by solving
\begin{equation}\label{eq:robust_meas}
\begin{split}
      &\tau_{A_1}=\min_{d\in \mathcal{Q}_5}\,\mathsf{T}_{A_1}\\
      \text{s.t.}\;&\tr\left(\frac{\mathbb{1}+Z_{A_1}}{2} \otimes \frac{\mathbb{1}+Z_{A_2}}{2}\otimes \frac{\mathbb{1}+Z_{A_3}}{2} \rho_{A_1A_2A_3}\right)=p^{\max}_3-\varepsilon_1,\\
      &\tr\left(\frac{\mathbb{1}+X_{A_1}}{2} \otimes \frac{\mathbb{1}+Z_{A_2}}{2} \rho_{A_1A_2A_3}\right)\le\varepsilon_2,\\
      &\tr\left(\frac{\mathbb{1}+X_{A_2}}{2}\otimes \frac{\mathbb{1}+Z_{A_3}}{2} \rho_{A_1A_2A_3}\right)\le\varepsilon_2,\\
      &\tr\left(\frac{\mathbb{1}+Z_{A_1}}{2} \otimes \frac{\mathbb{1}+X_{A_3}}{2} \rho_{A_1A_2A_3}\right)\le\varepsilon_2,\\
      &\tr\left(\frac{\mathbb{1}-X_{A_1}}{2} \otimes \frac{\mathbb{1}-X_{A_2}}{2}\otimes \frac{\mathbb{1}-X_{A_3}}{2} \rho_{A_1A_2A_3}\right)\le\varepsilon_2,
\end{split}
\end{equation}

Note that trivial measurements, $\frac{\mathbb{1}+(-1)^aA^j_1}{2}=\frac{\mathbb{1}_2}{2}$ for all $a,j$, already achieve $\mathsf{T}_{A_1}=0$. Hence, a nontrivial self-test requires $\tau_{A_1}>0$. The corresponding robustness analysis is shown in Fig.~\ref{fig:fidelity_measure}.

From Fig.~\ref{fig:robustness} we observe that, for $N=3$, the condition \eqref{tri} can tolerate at most an error of $\varepsilon_2=10^{-4}$. Within this regime, the fidelity satisfies $\mathcal{F}>\tfrac{1}{2}$ provided $\varepsilon_1 \lesssim 0.005$. For this range of $\varepsilon_1$, we also find that $\tau_{A_1}>0$. This robustness analysis guarantees that even when \eqref{tri} is only approximately satisfied, the state $\ket{\psi^H_N}$ and associated measurements can still be self-tested reliably. Moreover, because $\varepsilon_2$ is extremely small, its effect is negligible in practice and does not significantly impact the performance of the ``\emph{Key Generation}'' step.

 \section{Noisy Key Rate for the Tripartite Hardy Protocol} \label{security}

In a realistic implementation the Hardy conditions cannot be satisfied
exactly due to statistical fluctuations and experimental imperfections.
To analyze the robustness of the protocol, we consider a noisy version
of the tripartite Hardy correlations and study how these deviations
affect the achievable secret key rate. In particular, when the Hardy
zero-probability conditions are only approximately satisfied, the
perfect correlation between the parties’ inputs is lost and the honest
parties must perform information reconciliation. In this section we
quantify the resulting key rate of Protocol~1 in the presence of such
noise.


In the presence of noise, the tripartite Hardy scenario becomes,

\begin{equation}
\begin{split}
P(+1,+1,+1|A^0_1,A^0_2,A^0_3) &= p^{\max}_3-\varepsilon_1, \\
P_{AB}(+1,+1|A^1_1,A^0_2) &\leq \varepsilon_2, \\
P_{BC}(+1,+1|A^1_2,A^0_3) &\leq \varepsilon_2, \\
P_{AC}(+1,+1|A^0_1,A^1_3) &\leq \varepsilon_2, \\
P(-1,-1,-1|A^1_1,A^1_2,A^1_3) &\leq \varepsilon_2,
\end{split}
\end{equation}
where $\varepsilon_1,\varepsilon_2 \ge 0$ quantify imperfections. These deviation parameters $(\varepsilon_1,\varepsilon_2)$ therefore determine the operational region in which the protocol proceeds: as long as the observed correlations satisfy the noisy Hardy constraints within these tolerances, the eavesdropping test described in Step~3 is passed and the protocol can continue to key generation.

\subsection{Key Generation and Dropping Strategy}
The raw key is generated from rounds in which all three parties obtain
the outcome $+1$. From the Hardy constraints, such events ideally occur only when
\begin{equation}
(A^{k_1}_1,A^{k_2}_2,A^{k_3}_3) = (A^0_1,A^0_2,A^0_3) \quad \text{or} \quad (A^1_1,A^1_2,A^1_3).
\end{equation}

We assign the key bit as
\begin{equation}
K =
\begin{cases}
0 & \text{if } (A^{k_1}_1,A^{k_2}_2,A^{k_3}_3)=(A^0_1,A^0_2,A^0_3), \\
1 & \text{if } (A^{k_1}_1,A^{k_2}_2,A^{k_3}_3)=(A^1_1,A^1_2,A^1_3).
\end{cases}
\end{equation}

Since generally $P(+1,+1,+1|A^0_1,A^0_2,A^0_3) < P(+1,+1,+1|A^1_1,A^1_2,A^1_3)$, the key is biased. So, we apply a dropping strategy where $\mathsf{d}=\frac{P(+1,+1,+1|A^0_1,A^0_2,A^0_3)}{P(+1,+1,+1|A^1_1,A^1_2,A^1_3)}$ of the $(A^1_1,A^1_2,A^1_3)$ rounds is retained.

After dropping, the probability per experimental round that a key bit is generated is
\begin{equation}
\Omega =
\frac{1}{8} P(+1,+1,+1|A^0_1,A^0_2,A^0_3)
+
\frac{\mathsf{d}}{8} P(+1,+1,+1|A^1_1,A^1_2,A^1_3)
=
\frac{P(+1,+1,+1|A^0_1,A^0_2,A^0_3)}{4}.
\end{equation}

Under noise $\Omega$ becomes,
\begin{equation}
\Omega(\varepsilon_1)
=\frac{p^{\max}_3-\varepsilon_1}{4}.
\end{equation}

\subsection{Quantum Bit Error Rate}

In the ideal tripartite Hardy protocol, a key bit is generated only when
all three parties obtain the outcome $+1$, and this event occurs
exclusively for the input settings
$(A_1^0,A_2^0,A_3^0)$ or $(A_1^1,A_2^1,A_3^1)$.
In this ideal case the inputs are perfectly correlated and therefore
no errors arise.

In the presence of noise, however, the Hardy-zero conditions are no
longer strictly satisfied. Consequently, it becomes possible that the
outcome $(+1,+1,+1)$ occurs for other input combinations.
Such events lead to inconsistencies in the inferred key bit and
therefore contribute to the quantum bit error rate (QBER).

From the noisy Hardy constraints, each forbidden event is bounded by
$\varepsilon_2$. There are six such input combinations,
\[\{(A_1^0,A_2^0,A_3^1),(A_1^0,A_2^1,A_3^0),(A_1^1,A_2^0,A_3^0),(A_1^0,A_2^1,A_3^1),(A_1^1,A_2^1,A_3^0),(A_1^1,A_2^0,A_3^1)\},\]
for which the event $(+1,+1,+1)$ may occur. Since the input settings
are chosen uniformly, the total probability per experimental round
of such erroneous events is bounded by
\begin{equation}
P_{\mathrm{err}} \le \frac{6}{8}\varepsilon_2 .
\end{equation}

The QBER is defined as the ratio between the probability of erroneous
key-generating events and the total probability that a key bit is
generated. Note that $\Omega(\varepsilon_1)$ is the probability that a
round produces a key bit. Hence we obtain
\begin{equation}
QBER(\varepsilon_1,\varepsilon_2)
=
\frac{P_{\mathrm{err}}}
{\Omega(\varepsilon_1) + P_{\mathrm{err}} } .
\end{equation}

Using $\Omega(\varepsilon_1)=\frac{p_3^{\max}-\varepsilon_1}{4}$ and
the bound above on $P_{\mathrm{err}}$, the QBER can therefore be
bounded as
\begin{equation}
QBER(\varepsilon_1,\varepsilon_2)
\le
\frac{3\varepsilon_2}
{p_3^{\max}-\varepsilon_1+3\varepsilon_2}.
\end{equation}

This expression shows that the QBER grows linearly with the noise
parameter $\varepsilon_2$ and vanishes in the ideal Hardy limit
$(\varepsilon_1,\varepsilon_2)\to(0,0)$.





\subsection{Device-Independent Guessing Probability}

To quantify Eve’s information about the generated key, we adopt the
standard device-independent framework. We assume that the tripartite
state $\rho_{A_1A_2A_3}$ shared by $A_1$, $A_2$, and $A_3$ admits a
purification held by an adversary Eve. That is, the global state can
be written as a pure state $\ket{\Psi}_{A_1A_2A_3E}$, where Eve’s
system $E$ contains all side information compatible with the observed
statistics. No assumptions are made about the dimension or internal
structure of Eve’s system.

Eve attempts to guess the key bit produced in a given round. Recall
that the key is defined through the input settings that lead to the
event $(+1,+1,+1)$, namely $(A_1^0,A_2^0,A_3^0)$ corresponding to key
bit $0$, and $(A_1^1,A_2^1,A_3^1)$ corresponding to key bit $1$.
Eve therefore performs a binary measurement on her system, which we
represent by a dichotomic observable $E = E_0 - E_1$, where the
outcomes $+1$ and $-1$ correspond to her guesses of key bit $0$ and
$1$, respectively.

Eve’s strategy is successful if her outcome $+1$ coincides with the
honest parties obtaining $(+1,+1,+1)$ under the settings
$(A_1^0,A_2^0,A_3^0)$, or if her outcome $-1$ coincides with the
event $(+1,+1,+1)$ under $(A_1^1,A_2^1,A_3^1)$. Taking into account
the dropping factor $\mathsf{d}$ that balances the key distribution
between the two setting choices, the total joint success probability
per experimental round is given by
\begin{equation}
P_{\mathrm{joint}}
=
P(+1,+1,+1,E=+1|A_1^0,A_2^0,A_3^0)
+
\mathsf{d}\,P(+1,+1,+1,E=-1|A_1^1,A_2^1,A_3^1).
\end{equation}

Since a key bit is produced only with probability
$\Omega(\varepsilon_1)$, Eve’s relevant success probability is the
conditional probability that she correctly guesses the key bit given
that a key-generating event has occurred. This quantity defines the
device-independent guessing probability,
\begin{equation}
P_G(\varepsilon_1,\varepsilon_2)
=
\frac{P_{\mathrm{joint}}}{\Omega(\varepsilon_1)}.
\end{equation}

In the device-independent setting, $P_G(\varepsilon_1,\varepsilon_2)$
is upper bounded by solving a semidefinite program over the set of
quantum correlations compatible with the observed noisy Hardy
constraints. In our analysis this optimization is performed using the NPA hierarchy at level $3$. The NPA
relaxation allows us to compute an upper bound on Eve's optimal
guessing probability consistent with the observed behavior of the
devices.

The resulting bound on $P_G(\varepsilon_1,\varepsilon_2)$ determines
the conditional min-entropy of the key given Eve,
\begin{equation}
H_{\min}(A|E) = -\log_2 P_G(\varepsilon_1,\varepsilon_2).
\end{equation}

A smaller value of $P_G$ therefore corresponds to greater intrinsic
randomness of the generated key and leads to a higher achievable
secret key rate.

\subsection{Secret Key Rate}

Having bounded both Eve's information and the intrinsic error rate of
the protocol, we now determine the achievable secret key rate. In the
device-independent setting, the asymptotic secret key rate against
collective attacks is given by the Devetak--Winter bound
\cite{Devetak2005},
\begin{equation}
r \ge H(A|E) - \max\{H(A|B),H(A|C)\},
\end{equation}
where $H(A|E)$ quantifies Eve's uncertainty about Alice's key and the
second term represents the cost of classical error correction between
the honest parties.

In the device-independent framework, a convenient lower bound on
$H(A|E)$ is obtained from the conditional min-entropy, which is
directly related to Eve's optimal guessing probability,
\begin{equation}
H_{\min}(A|E) = -\log_2 P_G(\varepsilon_1,\varepsilon_2).
\end{equation}
Since $H(A|E) \ge H_{\min}(A|E)$, we obtain the bound
\begin{equation}
H(A|E) \ge -\log_2 P_G(\varepsilon_1,\varepsilon_2).
\end{equation}

Furthermore, under the assumption of symmetric noise, the correlations
between $A_1$ and the other parties are well characterized by the
quantum bit error rate. In this case the conditional entropy can be
approximated by the binary Shannon entropy,
\begin{equation}
H(A|B) \approx H(A|C) \approx h(QBER),
\end{equation}
where $h(x) = -x\log_2 x - (1-x)\log_2(1-x)$.

Since a key bit is produced only with probability
$\Omega(\varepsilon_1)$, the secret key rate per experimental round is
\begin{equation}
K(\varepsilon_1,\varepsilon_2)
=
\Omega(\varepsilon_1)
\left[
-\log_2 P_G(\varepsilon_1,\varepsilon_2)
-
h\!\left(QBER(\varepsilon_1,\varepsilon_2)\right)
\right].
\end{equation}

Substituting $\Omega(\varepsilon_1)=\frac{p_3^{\max}-\varepsilon_1}{4}$,
we obtain the final expression for the noisy tripartite Hardy key rate
\begin{equation}
K(\varepsilon_1,\varepsilon_2)
=
\frac{p_3^{\max}-\varepsilon_1}{4}
\left[
-\log_2 P_G(\varepsilon_1,\varepsilon_2)
-
h\!\left(QBER(\varepsilon_1,\varepsilon_2)\right)
\right].
\end{equation}

This expression shows that a positive secret key rate can be obtained
whenever Eve's uncertainty about the key, quantified by the term
$-\log_2 P_G(\varepsilon_1,\varepsilon_2)$, exceeds the information
revealed during error correction, quantified by
$h(QBER(\varepsilon_1,\varepsilon_2))$.

\begin{figure}
    \centering
    \includegraphics[width=0.5\linewidth]{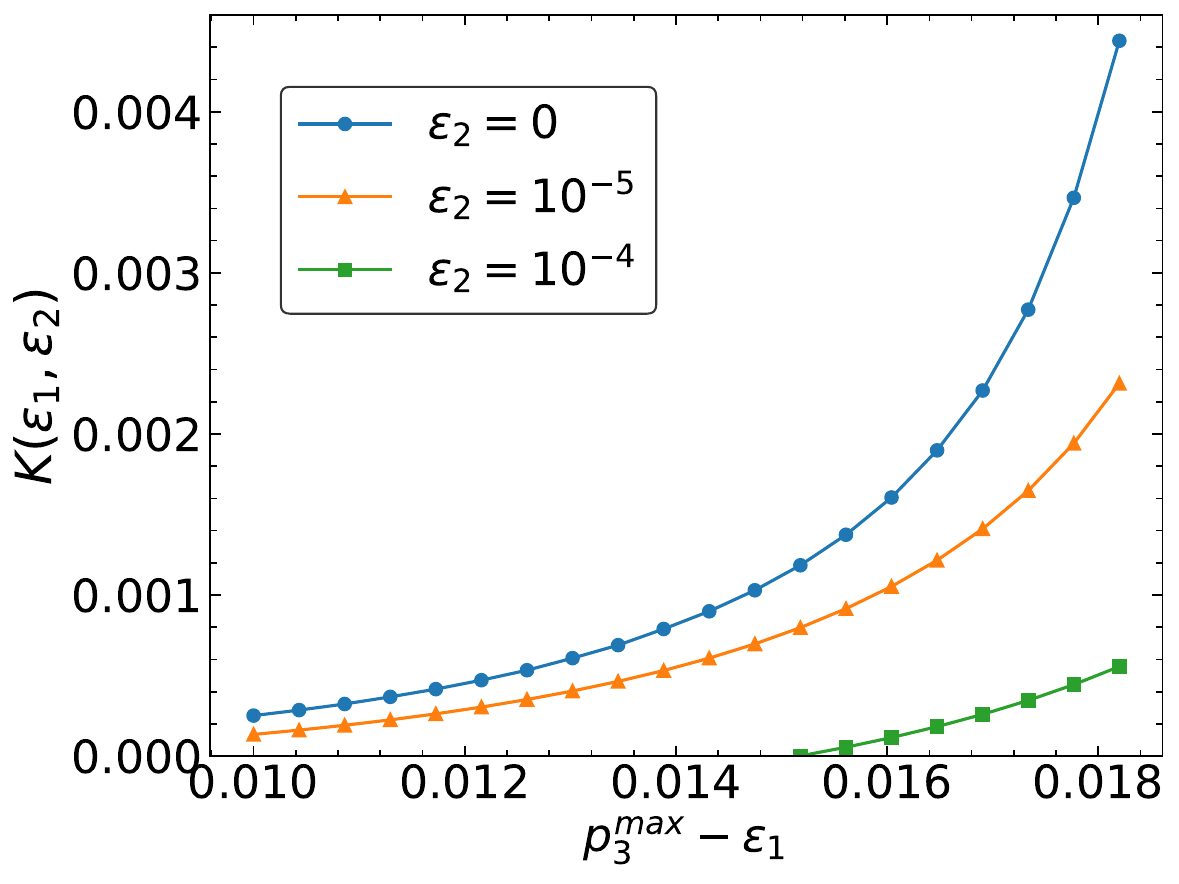}
    \caption{Secret key rate of the noisy tripartite Hardy protocol as a
function of the Hardy success probability
$P(+1,+1,+1|A^0_1,A^0_2,A^0_3)=p^{\max}_3-\varepsilon_1$,
for several values of the noise parameter $\varepsilon_2$.
The curves show how the achievable key rate decreases as the Hardy
correlations deviate from their ideal values.
A positive key rate is obtained only within a finite noise regime;
beyond the point where the curves reach zero, the protocol must abort.}
    \label{fig:noisykeyrate}
\end{figure}

The dependence of the secret key rate $K(\varepsilon_1,\varepsilon_2)$ on the deviation parameter $\varepsilon_1$ for different values of $\varepsilon_2$ is illustrated in Fig.~\ref{fig:noisykeyrate}. The plot shows that the key rate remains positive only within a limited noise regime, thereby providing an operational threshold for acceptable deviations from the ideal Hardy correlations. The parameters $\varepsilon_1$ and $\varepsilon_2$ quantify the admissible deviations from the ideal Hardy relations. Our analysis shows that the protocol yields a positive key rate only when these deviations remain within tolerances $\varepsilon_1 \lesssim 0.002$ and $\varepsilon_2 \le 10^{-4}$. For larger deviations the key rate becomes zero, and the protocol must abort.





\section{Robustness against imperfect randomness} 
\label{sec:biased}

In this section, we compare the robustness of our approach against compromised random number generators (RNGs) with that of DICKA protocols based on the tripartite Holz inequality and the tripartite Parity-CHSH inequality. In all cases, we assume that the observed statistics are those expected from perfect quantum states and measurement devices.

In standard CKA analyses, one typically assumes that the source of randomness is ideal: the measurement settings are chosen independently and identically distributed with uniform probabilities. This corresponds to the distribution
\begin{equation}
\label{idealDistribution}
\mathcal{P}_{ideal}(A_1, A_2, A_3) =
\begin{cases}
P(A^0_1,A^0_2,A^0_3) = p_{A_1} p_{A_2} p_{A_3}, \\
P(A^0_1,A^1_2,A^0_3) = p_{A_1} (1 - p_{A_2}) p_{A_3}, \\
P(A^1_1,A^0_2,A^0_3) = (1 - p_{A_1}) p_{A_2} p_{A_3}, \\
P(A^1_1,A^1_2,A^0_3) = (1 - p_{A_1})(1 - p_{A_2}) p_{A_3}, \\
P(A^0_1,A^0_2,A^1_3) = p_{A_1} p_{A_2} (1 - p_{A_3}), \\
P(A^0_1,A^1_2,A^1_3) = p_{A_1} (1 - p_{A_2})(1 - p_{A_3}), \\
P(A^1_1,A^0_2,A^1_3) = (1 - p_{A_1}) p_{A_2}(1 - p_{A_3}), \\
P(A^1_1,A^1_2,A^1_3) = (1 - p_{A_1})(1 - p_{A_2})(1 - p_{A_3}),
\end{cases}
\end{equation}
with $p_{A_1} = p_{A_2} = p_{A_3} = \tfrac{1}{2}$ in the uniform case.

Here, however, we focus on the more realistic scenario where the average distribution still coincides with ~\eqref{idealDistribution}, but in individual runs the RNGs may be biased in a way known to the eavesdropper. For simplicity, we model this bias by shifting $p_{A_1}$, $p_{A_2}$, and $p_{A_3}$ by $\pm \varepsilon$, which yields eight possible biased distributions $\{\mathcal{P}_{b,i}\}_{i=1,\ldots,8}$. The average distribution is recovered only if all eight biased cases occur equally often.

In the tripartite CKA Hardy protocol, $A_1$, $A_2$, and $A_3$ can choose their measurement setting with non-uniform probability distribution such that $P(A_1^0,A_2^0, A_3^0)P(+1,+1,+1 | A_1^0,A_2^0, A_3^0)=P(A_1^1,A_2^1, A_3^1)P(+1,+1,+1 | A_1^1,A_2^1, A_3^1)$. So that $A_1$ does not have to drop some of the runs where $+1$ outcome of the measurement $A_1^1$ gets clicked. In that case, all the parties choose measurement $A^0_i$ with probability $r_H\approx 0.6478024$. We are considering an Eve that can control the RNG in such a way that the parties on average sees their input random distribution to follow $\{r_H,1-r_H\}$ in Hardy scenario and $\{\frac{1}{2},\frac{1}{2}\}$ in tripartite Holz and Parity-CHSH scenario.

This assumption has significant implications. If one only observes the averaged distribution, then during runs with a particular biased distribution $\mathcal{P}_{b,i}$, the actual conditional probabilities are systematically under or over estimated:
\begin{equation}
\label{underOverEstimated}
\begin{split}
&P_{obs,i}(a_1,a_2,a_3|A_1,A_2,A_3) \\
&= P_{actual}(a_1,a_2,a_3|A_1,A_2,A_3)
\frac{P_{b,i}(A_1,A_2,A_3)}{P_{ideal}(A_1,A_2,A_3)}.
\end{split}
\end{equation}

To illustrate, consider the noiseless Hardy-type correlations with,$P(+1,+1,+1|A_1^0,A_2^0,A_3^0) = q$, and $P(+1,+1,+1|A_1^1,A_2^1,A_3^1) = \tilde{q}$.

Under biased distributions $\mathcal{P}_{b,i}$, the probability that the key bit equals $0$ is
\begin{equation*}
P_{i,\text{key}=0} =
\frac{q P_{b,i}(A^0_1,A^0_2,A^0_3)}{q P_{b,i}(A^0_1,A^0_2,A^0_3) + \tilde{q} P_{b,i}(A^1_1,A^1_2,A^1_3)}.
\end{equation*}
The eavesdropper maximizes her advantage by guessing the more probable key value, giving
\begin{equation*}
P_{guess,i} = \max\big(P_{i,\text{key}=0},  1 - P_{i,\text{key}=0}\big).
\end{equation*}
Averaging uniformly over all eight biased cases yields the overall guessing probability:
\begin{equation}
\mathscr{P}_{guess}=\frac{1}{8} \sum_{i=1}^8 P_{guess,i}.
\end{equation}

\begin{figure}[htbp]
\centering
\includegraphics[width=0.48\textwidth]{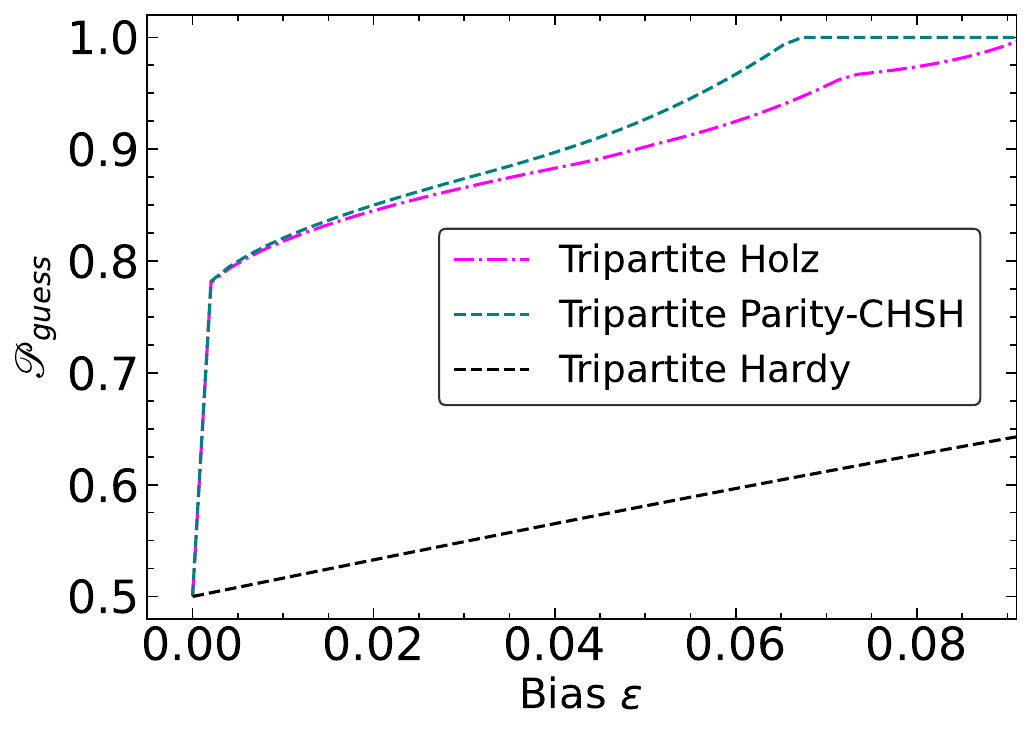}
\caption{Comparison of the eavesdropper’s guessing probability in CKA protocols based on the Hardy paradox, the tripartite Holz inequality, and the tripartite Parity-CHSH inequality under biased setting distributions.}
\label{fig:baised}
\end{figure}

The resulting guessing probabilities as a function of $\varepsilon$ are shown in Fig.~\ref{fig:baised}.

For comparison, we apply the method of ~\cite{navascues2007bounding,navascues2008convergent} to evaluate the guessing probability for protocols based on the tripartite Holz and Parity-CHSH inequalities. Specifically, we compute the bound on $A_1$’s outcome-guessing probability implied by a maximal violation of the corresponding inequality ($\tfrac{3}{2}$ and $\sqrt{2}$, respectively) under biased setting distributions. The relevant expressions are
\begin{equation}\label{holz}
\begin{aligned}
\beta_H=8 \Big(& \tfrac{P_{b,i}(A_1^1 ,A_2^0 ,A_3^0)\langle A_1^1 A_2^0 A_3^0\rangle}{4}+
\tfrac{P_{b,i}(A_1^1, A_2^0, A_3^1)\langle A_1^1 A_2^0 A_3^1\rangle}{4} 
  + \tfrac{P_{b,i}(A_1^1, A_2^1, A_3^0)\langle A_1^1 A_2^1 A_3^0\rangle}{4}
+\tfrac{P_{b,i}(A_1^1, A_2^1, A_3^1)\langle A_1^1 A_2^1 A_3^1\rangle}{4}\\ 
  -&\tfrac{P_{b,i}(A_1^0, A_2^0, A_3^0)\langle A_1^0 A_2^0\rangle}{2}
+\tfrac{P_{b,i}(A_1^0, A_2^1, A_3^0)\langle A_1^0 A_2^1\rangle}{2} 
  -\tfrac{P_{b,i}(A_1^0, A_2^0, A_3^0)\langle A_1^0 A_3^0\rangle}{2}
+\tfrac{P_{b,i}(A_1^0, A_2^0, A_3^1)\langle A_1^0 A_3^1\rangle}{2}\\ 
  -&\tfrac{P_{b,i}(A_1^0, A_2^0 A_3^0)\langle A_2^0 A_3^0\rangle}{4}
+\tfrac{P_{b,i}(A_1^0, A_2^0 A_3^1)\langle A_2^0 A_3^1\rangle}{4} 
  + \tfrac{P_{b,i}(A_1^0, A_2^1 A_3^0)\langle A_2^1 A_3^0\rangle}{4}
- \tfrac{P_{b,i}(A_1^0, A_2^1 A_3^1)\langle A_2^1 A_3^1\rangle}{4} \Big),
  \end{aligned}
  \end{equation}
  and
  
  \begin{equation}\label{paritychsh}
  \begin{aligned}
  \beta_{PC}=8 \Big(\frac{1}{2} (P_{b,i}(A_1^1, A_2^0, A_3^0)\langle A_1^1 A_2^0 A_3^0\rangle
- P_{b,i}(A_1^1, A_2^1, A_3^0)\langle A_1^1 A_2^1 A_3^0\rangle
  + P_{b,i}(A_1^0, A_2^0, A_3^0)\langle A_1^0 A_2^0\rangle
+P_{b,i}(A_1^0, A_2^1, A_3^0)\langle A_1^0 A_2^1\rangle) \Big),
  \end{aligned}
  \end{equation}
  where the factor $8$ corresponds to the inverse of the uniform probability of each setting pair in the unbiased case. As before, the final guessing probability is obtained by averaging over all eight biased distributions. The optimization problem has been implemented in PYTHON using CVXPY~\cite{gitmishra}.

The results, summarized in Fig.~\ref{fig:baised}, reveal that the Holz and Parity-CHSH inequality fail (yielding zero key rate) already for biases of about $\epsilon \approx 0.09$, the Hardy-based protocol continues to provide positive key rates well beyond this threshold. This demonstrates the superior robustness of our approach to imperfections in the sources of randomness.

\section{Conclusion}\label{sec:conclu}
In this   article, we present a novel device-independent conference key agreement (DICKA) protocol tailored for arbitrary $N$ parties, harnessing the multipartite Hardy paradox. Unlike conventional approaches, our protocol generates the shared secret key directly from measurement settings under maximal violation, bypassing the need to rely solely on measurement outcomes. This design choice simplifies the key extraction process and reduces dependence on maximally entangled state, a resource notoriously difficult to maintain in experimental settings due to its fragility. By employing non-maximally entangled genuine multipartite states,   which demand less entanglement as a resource (see Table \ref{tab:entanglement_measures}) than maximally entangled states $GHZ$, our method achieves positive key rates while enhancing practicality. A distinguishing feature of our approach is that any two participants can achieve pairwise key rates that substantially exceed the collective $N$-party rate.

\begin{table}[htpb!]
    \begin{center}
    \begin{tabular}{|c|c|c||c|c|}
    \hline
     & GHZ & $\ket{\psi^{H}_3}$ & GHZ$_4$ & $\ket{\psi^{H}_4}$ \\ \hline
    Concurrence & 1 & 0.54 & 1 &  0.37 \\ \hline
    Negativity & 0.5 & 0.27 & 0.5 &  0.19 \\ \hline
    Log Negativity & 1 & 0.62 & 1 & 0.46  \\ \hline
    Entanglement Entropy & 1 & 0.4 & 1 &  0.22  \\ \hline
    Rényi Entanglement Entropy & 1 & 0.62 & 1 & 0.46  \\ \hline
    \end{tabular}
    \caption{  Comparison of different entanglement monotones in the $1$ versus $2$ bipartition for three-qubit states and the $1$ versus $3$ bipartition for four-qubit states. In all cases, the Hardy states $\ket{\psi^{H}_3}$ and $\ket{\psi^{H}_4}$ exhibit strictly lower values than the corresponding GHZ states. This indicates that Hardy states require less entanglement as a resource, making them easier to generate in practice, compared to GHZ states.}
    \label{tab:entanglement_measures}  
    \end{center}
\end{table}

 Beyond these core findings, our analysis demonstrates that a direct multipartite approach provides distinct operational advantages over a sequence of bipartite DIQKD rounds. By utilizing a single global resource, the protocol minimizes the entanglement distribution phases and reduces the classical communication overhead required for error reconciliation. While Protocol 2 is primarily intended as a theoretical demonstration of the ``intrinsic power'' of Hardy correlations to elevate key rates in the ideal limit, it establishes a high-value benchmark for future architectures . Although measurement input-based key distribution protocol addresses many practical challenges, the current protocol suffers from diminishing key rate as the number of parties becomes very large. Devising a measurement input-based protocol which ensures a sufficiently large key rate for arbitrary large number of parties can be a potential direction for future research. 

On the other hand, the maximal violation of our multipartite Hardy paradox is highly sensitive to external noise. A compelling direction would be to explore new multipartite paradoxes, akin to Hardy’s, that retain strong nonlocal properties while offering greater resilience to noise. Such paradoxes could potentially sustain or even enhance key rates across larger networks, striking a balance between security and practicality. Developing these advancements could elevate device-independent multiparty cryptography from a theoretical innovation to a cornerstone of secure quantum communication, bridging the gap between idealized models and operational reality.

\section*{Acknowledgments}
 We are grateful for stimulating discussions with Arup Roy, Subhendu Bikash Ghosh, Snehasish Roy Chowdhury, Shayeef Murshid, Utkarsh Sahai, and Ramprasad Sarkar. R.A. acknowledges financial support from the Council of Scientific and Industrial Research (CSIR), Government of India, under File No. 09/0093(19292)/2024-EMR-I. M.M acknowledges support from the Science and Engineering Research Board (SERB), Department of Science and Technology (DST), Government of India, under File No. EEQ/2023/000164 and the Information Security Education and Awareness (ISEA) Project Phase-III initiatives of the Ministry of Electronics and Information Technology (MeitY) under Grant No. F.No. L-14017/1/2022-HRD.

\bibliography{qkdref}

\end{document}